# Decoupling minimal surface metamaterial properties through multi-material hyperbolic tilings

Sebastien J.P. Callens[*,1], Christoph H. Arns[2], Alina Kuliesh[1], Amir A. Zadpoor[1]

[1] *Department of Biomechanical Engineering, TU Delft, Mekelweg 2, 2628CD Delft, The Netherlands*

[2] *School of Minerals and Energy Resources Engineering, The University of New South Wales, Sydney NSW 2052, Australia*

**ABSTRACT**

Rapid advances in additive manufacturing over the past decade have kindled widespread interest in the rational design of metamaterials with unique properties. However, many applications require multi-physics metamaterials, where multiple properties are simultaneously optimized. This is challenging, since different properties, such as mechanical and mass transport properties, typically impose competing requirements on the nano-/micro-/meso-architecture of metamaterials. Here, we propose a parametric metamaterial design strategy that enables independent tuning of the effective permeability and elastic properties. We apply hyperbolic tiling theory to devise simple templates based on which triply periodic minimal surfaces (TPMS) are partitioned into hard and soft regions. Through computational analyses, we demonstrate how the decoration of hard, soft, and void phases within the TPMS substantially enhances their permeability-elasticity property space and offers high tunability in the elastic properties and anisotropy, at constant permeability. We also show that this permeability-elasticity balance is well captured using simple scaling laws. We then proceed to demonstrate the proposed concept through multi-material additive manufacturing of representative specimens. Our approach, which is generalizable to other designs, offers a route towards multi-physics metamaterials that need to simultaneously carry a load and enable mass transport, such as load-bearing heat exchangers or architected tissue-substituting meta-biomaterials.

[*] Corresponding author: s.j.p.callens@tudelft.nl.

# 1. INTRODUCTION

The fundamental paradigm of metamaterials is that their macroscale properties are largely driven by their nano-, micro- or mesoscale architecture. This intimate structure-property connection has been leveraged to develop metamaterials with unique, unusual, and extreme acoustic [1], photonic [2], or mechanical properties [3]. Historically, most types of metamaterial architectures have been based on periodic arrangements of struts, often inspired by crystallographic lattices [4]. In search for higher mass-specific mechanical properties, periodic plate-lattices have been proposed, capable of storing strain energy more efficiently [5]. More recently, smooth shell-based lattices have also attracted great interest, since these architectures are devoid of the stress concentrations that are inherent at the intersections of strut- or plate-lattices, and since their intrinsically curved morphology endows them with high specific stiffness and attractive energy absorption behavior [6]. Among the shell-based metamaterials, those derived from triply periodic minimal surfaces (TPMS) have most widely been studied [7]. These are bicontinuous, infinitely-extending, saddle-shaped surfaces that locally minimize area and have the defining characteristic of zero mean curvature ($H = 0$) at every point along the surface [8]. The widespread interest in TPMS-based structures has partly been fueled by their intriguing mathematical foundation and their widespread observations in spontaneously-assembled natural systems [9], but is also due to their attractive and extremal physical properties [10].

Irrespective of the architecture type, the central challenge in metamaterial design is to optimize the material geometry to attain the desired macroscale physical properties. In the case of multi-physics metamaterials, however, several properties are targeted simultaneously. It turns out that optimizing the geometry for one property often leads to a decrease in the performance with respect to the others. This is exemplified in architected tissue scaffolds, or "meta-biomaterials" [11], where the material geometry has conflicting effects on the mechanical and mass transport functionality [12]: increasing the mechanical properties, by increasing the relative density of the metamaterial [13], generally results in a decreased fluid permeability. A notable example where the decoupling of these properties is somewhat possible is a pentamode metamaterial, consisting of spindle-shaped struts that meet at relatively weak nodes [14]. The mechanical properties of these materials mainly depend on the node geometry and not on the overall relative density, offering the ability to partially tune the permeability independently of the mechanical properties [15]. However, this ability is limited, because the upper bound on permeability is constrained by the desired mechanical properties, and because pentamode metamaterials inherently rely on highly-specific strut-based architectures. As an alternative, one could consider scaling of the unit cells. A uniform scaling of the unit cell length by a factor $l$ does not affect the elastic properties of the lattice, while the absolute permeability scales with $l^2$. This implies that one could tune both properties somewhat independently, merely by scaling the structure [12c]. However, in many applications of multifunctional metamaterials, the unit cell size is not a parameter that could freely be altered, at least not without affecting other relevant properties or violating the requirements of the applied manufacturing processes. In meta-biomaterials, for example, the pore size should remain within experimentally determined bounds to promote tissue regeneration [16]. Moreover, scaling of the unit cell size would also affect properties, such as overall cell attachment or biodegradation behavior, which are dependent on the specific surface area and scale with $l^{-1}$ [12c]. Furthermore, the resolution of the additive manufacturing process or the desired number of unit cells to obtain sufficiently homogenized behavior could impose additional constraints on unit cell scaling.

An attractive and more potent strategy to unlock a larger metamaterial design space is to spatially distribute multiple materials with widely different properties, instead of architecting only a single material. This approach has only recently become possible, owing to advances in multi-material additive manufacturing, and has enabled the design of mechanical metamaterials and composites with exotic deformation modes and tunable Poisson's ratios [17]. Here, we leverage the multi-material strategy to develop periodic metamaterials with independently tunable properties. Specifically, we propose a strategy to parametrically design biphasic, TPMS-based architectures that interpolate between strut- and shell-lattices. Our design strategy builds upon the inherent hyperbolic symmetries of TPMS, offering a robust approach to tailor unit cell geometry and the spatial distribution of the different materials. This enables us to decouple the mechanical and mass transport properties to an extent that is not possible in uniphasic metamaterials. Using computational homogenization, we determine the effective elastic properties and anisotropy of a wide range of structures as a function of the unit cell geometry and material choice. Moreover, we quantify the intrinsic permeability (*i.e.* normalized by unit cell length) of the metamaterials using computational fluid dynamics (CFD). Our results confirm that our parametric design strategy and the combination of two different materials significantly expands the achievable space of multi-physics properties and greatly enhances the ability to independently tune the permeability and elastic properties. Additionally, we demonstrate the proposed concept by additively manufacturing and mechanically testing metamaterials that combine hard and soft polymers. While we focus here on two types of TPMS, this concept is directly extendable to other types of TPMS and could also be generalized to other types of shell-lattices, even those of a stochastic nature. Ultimately, this approach of spatially decorating shell-based metamaterials with multiple materials could be useful in many applications where mechanical and mass transport properties are both important, such as load-bearing heat exchangers, noise-abating permeable airfoils, or architected tissue scaffolds.

## 2. RESULTS

### 2.1. Triply periodic networks from hyperbolic tilings

The foundation of our design approach is the intimate connection between TPMS and the hyperbolic geometry: the geometry of saddle shapes (with negative Gaussian curvature). Every TPMS can be constructed from a single, fundamental patch that is symmetrically patterned throughout 3D space. This repeating pattern corresponds to a triangular tiling on the hyperbolic plane $\mathbb{H}^2$. Essentially, this implies that a portion of 2D hyperbolic space ($\mathbb{H}^2$) can be projected onto a TPMS (with minor distortions) in 3D Euclidean space ($\mathbb{E}^3$), in a manner similar to how a portion of the 2D Euclidean plane ($\mathbb{E}^2$) can be embedded in $\mathbb{E}^3$ by wrapping it on a cylinder [18]. Here, we focus on two well-known TPMS of cubic symmetry, namely the P (primitive) and G (gyroid) surfaces. Both of these surfaces belong to the same TPMS family – they are related through the so-called Bonnet transformation – and they can both be derived from the same hyperbolic tiling. This hyperbolic tiling is called the *246 tiling (using orbifold notation), and consists of a repeating triangular patch with angles $\pi/2$, $\pi/4$, and $\pi/6$ (Figure 1a).

The remarkable connection between the hyperbolic plane and the TPMS enables the creation of a vast set of convoluted 3D networks, six of which are used as a template in this study. By decorating $\mathbb{H}^2$ with a periodic line pattern, *i.e.* a tiling that is a subgroup of the *246 tiling, and by wrapping that line pattern onto the P or G minimal surface, a periodic three-dimensional network, or surface reticulation, is obtained [18-19]. Here, we consider three different hyperbolic tilings that give rise to six periodic networks, though many other tilings are available to generate different networks [19c] (Figure 1b-d). The three hyperbolic tilings are obtained by drawing lines along one of the three edges of the fundamental patch. For example, the "26" tiling (Figure 1b) is obtained by connecting the edges between the $\pi/2$ and $\pi/6$ angles of all triangular patches. All of the six surface reticulations that we design are topologically equivalent to 3D networks that are known in reticular chemistry [20]. For example, the $P_{24}$ and $G_{46}$ surface reticulations correspond to the crystal network structures of the minerals sodalite and garnet, respectively. Notably, the same hyperbolic tiling wrapped onto the P or G surface results in substantially different networks from a topological perspective. For example, the "26" tiling (Figure 1b) on the P surface generates a network with a primitive cubic topology of genus 3 (per unit cell), while the same tiling on the G surface generates a much more complex network with genus 17, even though the P and G surfaces themselves are both of genus 3. It is interesting to note that the $P_{46}$ network consists entirely out of (Euclidean) straight lines, which is not the case for the other networks (Figure 1d). In fact, only a specific subset of TPMS, the so-called spanning minimal surfaces [21], have embedded straight lines, a property that is not shared by the G surface.

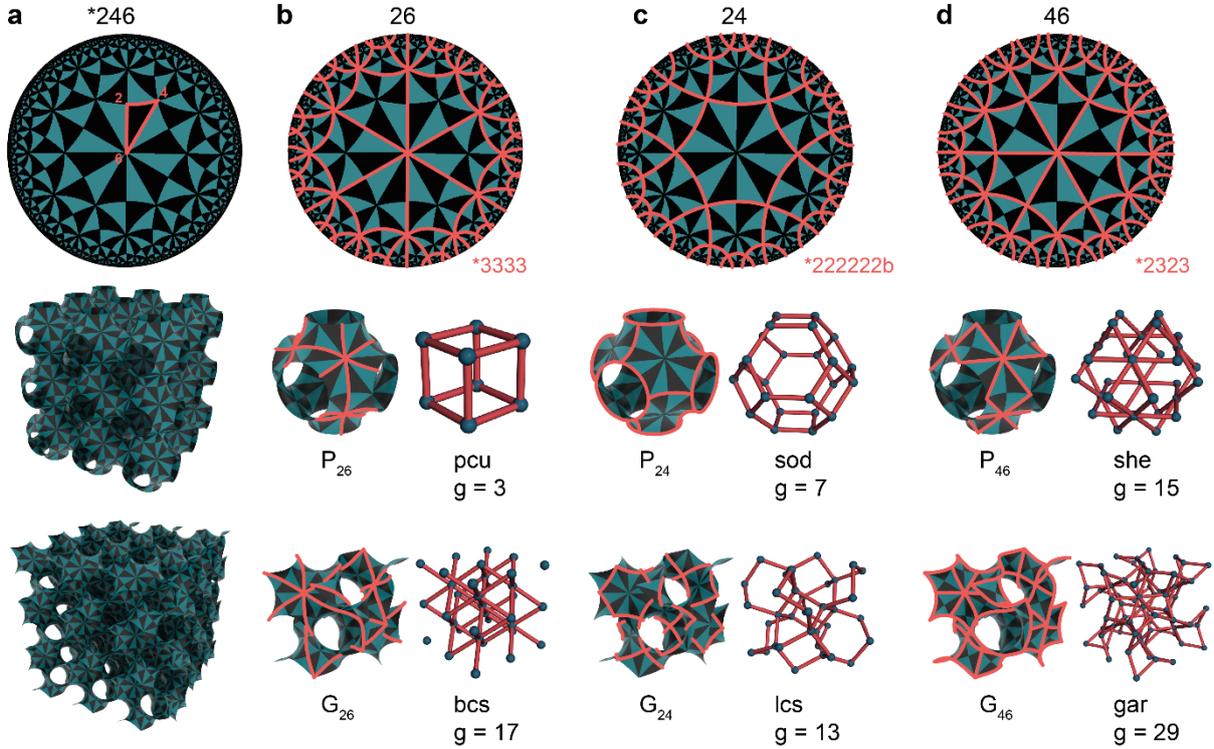

**Figure 1: Hyperbolic tilings projected onto TPMS.** a) The *246 hyperbolic tiling shown in the Poincaré disk model, with the fundamental triangular patch highlighted in red (top). The same hyperbolic tiling projected onto 27 translational unit cells of the P (middle) and G (bottom) surfaces. b) The "26" tiling in the Poincaré disk model (top) that results in two distinct networks on the P (middle) and G (bottom) surfaces. The orbifold naming of the tiling is indicated with red digits (*3333) and the RCSR naming [20] of the 3D net topology is indicated with three letters (pcu and bcs). We show both the network by its embedding in the TPMS (left) as well as its canonical form (right) [19c], which are topologically equivalent. The genus of the network (per unit cell) is indicated by g. c-d) Analogous to b), but now for the "24" and "46" hyperbolic tilings.

## 2.2. Parametric design of biphasic strut-shell metamaterials

The realization of 3D networks embedded in the P and G surfaces is the starting point of our strategy to parametrically design biphasic metamaterials. Essentially, our approach consists of "widening" these embedded networks to a desired degree, in order to form skeleton-like decorations on the TPMS that are templates to rationally distribute hard and soft phases (Figure 2). To construct the decorated translational unit cells of the P and G surfaces, we used the formal Enneper-Weierstrass parametrization [22]. This parametrization maps an integration domain in the complex plane $\mathbb{C}^2$ to the fundamental patch in $\mathbb{E}^3$, which is then symmetrically patterned to form the translational unit cell (Materials & Methods). The simplicity of the integration domain enabled us to easily label portions of it with either a hard or a soft phase. This labeling is transferred to the unit cell through the Enneper-Weierstrass mapping. We based the labeling of the integration domain on the previously-described hyperbolic tilings: the domain is subdivided into different regions through lines that are parallel to one of the three domain edges (Figure 2a). This subdivision is parametrized by two offset parameters $\phi_h \in [0,1]$ and $\phi_s \in [0,1]$, which respectively control the amount of hard and soft phases and are defined such that $\phi_h + \phi_s \leq 1$. The default scenario is to set $\phi_s = 1 - \phi_h$ and vary the offset parameter of the hard phase. When $\phi_h = 0$ or $\phi_h = 1$, the unit cell entirely consists of a soft or a hard phase, respectively. Any intermediate value of $\phi_h$ results in a biphasic partitioning of the unit cell, whereby the hard phase interpolates between predominantly strut-like or shell-like architectures (Figure 2a). It is also possible to decrease the offset parameter of the soft phase such that $\phi_h + \phi_s < 1$. In this case, not all points in the integration domain are utilized in the Enneper-Weierstrass mapping, and an incomplete fundamental patch is obtained. This results in a unit cell with additional openings as opposed to the traditional P or G morphology (Figure 2b).

We converted the zero-thickness surfaces into solid metamaterial unit cells by bidirectionally thickening the surface in the normal direction by a fraction of the bounding box length (Figure 2b). Using the three hyperbolic tilings shown in Figure 1, six distinct biphasic metamaterial morphologies could be generated with tunable amounts of hard and soft materials (by varying $\phi_s$ and $\phi_h$). We termed the designs with $\phi_s + \phi_h = 1$ "full" structures, and the designs with $\phi_s + \phi_h < 1$ "skeleton" structures (Figure 2b). The full biphasic designs with $\phi_s = 1 - \phi_h$ shown in Figure 2c and Figure 2d all have the same overall morphology, *i.e.* that of the standard P or G surface, yet exhibit widely different material decorations. It is important to observe that the hard phase always forms a triply connected structure, while the soft phase consists of isolated inclusions. For sufficiently small values of $\phi_h$, the hard phase essentially forms a strut-like skeleton that reinforces the predominantly soft-phased unit cell. For the larger values of $\phi_h$, the area of the soft inclusions diminishes and the hard-phased skeleton approaches the shell-like morphology of the original unit cell.

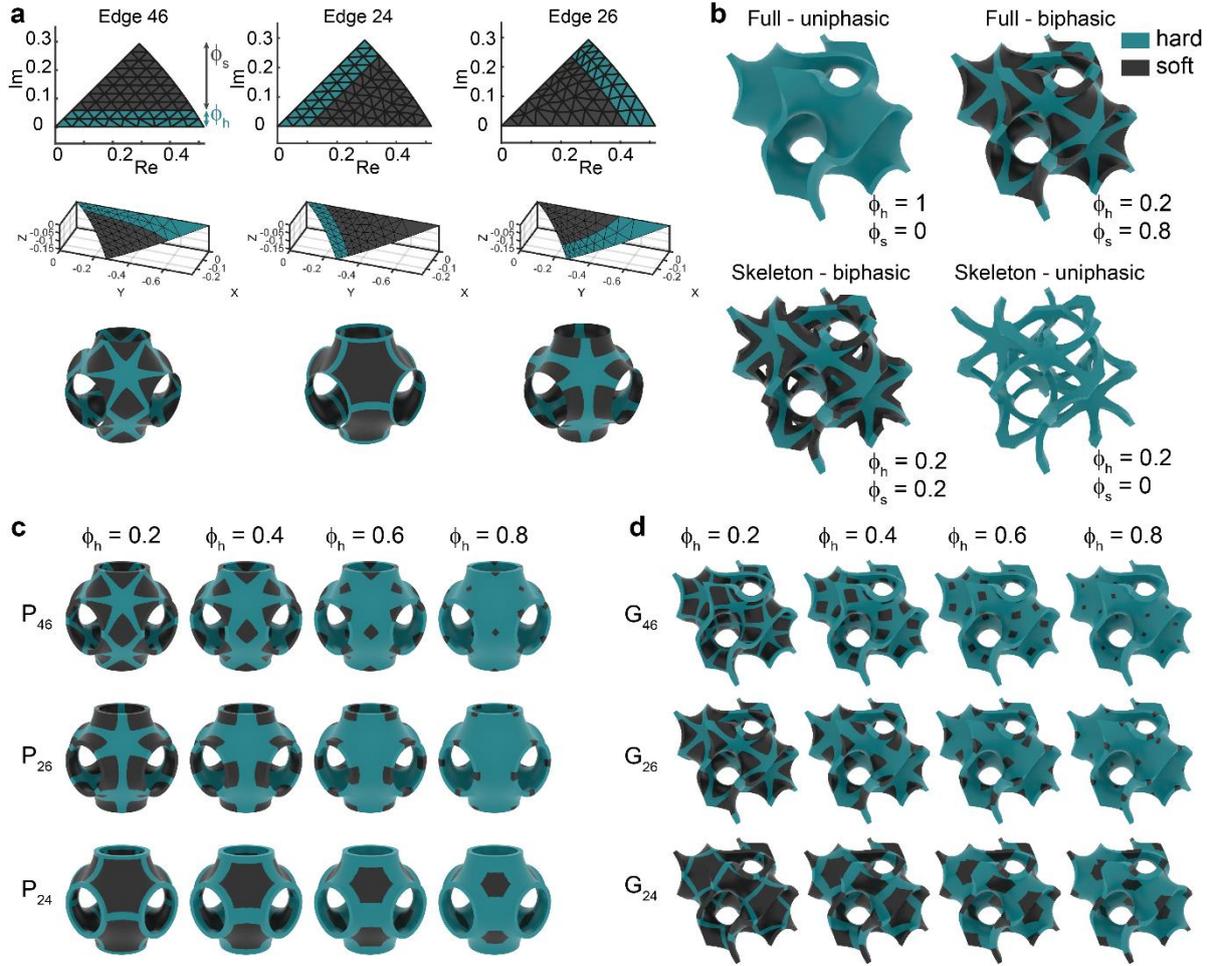

**Figure 2: Parametric design approach for biphasic TPMS.** a) Top row: the integration domain in the complex plane $\mathbb{C}^2$ that is parametrically partitioned into hard and soft regions for the "46", "24", and "26" designs. Middle row: the integration domain is mapped to $\mathbb{E}^3$ through the Enneper-Weierstrass parametrization, resulting in saddle-shaped fundamental patch with a biphasic partition. Bottom row: The translational unit cell of the P surface obtained through symmetry operations on the fundamental patches of the middle row. b) The different types of uni- and biphasic unit cell designs derived from the G surface. c-d) Metamaterial unit cells for the three P and G designs, respectively. In all cases, $\phi_s = 1$ and $\phi_h$ is varied between 0.2 and 0.8.

## 2.3. Morphology & mass transport properties

The defining characteristic of TPMS is their specific curvature profile: they are defined as surfaces with zero mean curvature ($H = 0$) and negative or vanishing Gaussian curvature ($K \leq 0$). Hence, TPMS are saddle-shaped everywhere, except at some isolated points where the surface is locally flat ($K = 0$). The specific curvature characteristic is part of the reason why TPMS have attracted interest as templates for tissue engineering scaffolds, since surface curvature is known to control the organization and dynamics of tissues and cells [23]. We quantified the curvature distributions of the P and G unit cells, as well as that of their skeletonized variants that are obtained at $\phi_h = 0.2$ and $\phi_s = 0$ (Figure 3a). We found that the "46" and "26" skeletonized designs of the P and G surfaces are, on average, less intrinsically curved than the "24" designs. Indeed, the $P_{46}$, $P_{26}$, $G_{46}$, and $G_{26}$ designs all maintain the locally flat regions in the skeletonized representations. These flat regions are connected through weakly or strongly curved ribbons in the "26" and "46" designs, respectively. In the $P_{24}$ and $G_{24}$ designs, however, the flat regions are absent and the skeletonized representation consists entirely out of highly curved ribbons (Figure 3a).

As mentioned before, the surfaces are converted to sheet-solids by offsetting the surface in both normal directions by a desired amount. This thickening operation, combined with variations in $\phi_s$ and $\phi_h$, enabled us to achieve a wide range of metamaterial volume fractions $\rho$. Here, $\rho = V_{solid}/L^3$, where $V_{solid}$ is the volume of the solid material and $L$ is the length of the cubic bounding box. We quantified the scaling of $\rho_h$, *i.e.* the volume fraction of the hard phase, with respect to the offset parameter $\phi_h$, finding that the "26" and "46" designs follow the same nonlinear scaling law in both the P (Figure 3b) and G (Figure 3e) surfaces. In the "24" designs, an almost linear relation between $\phi_h$ and $\rho_h$ is observed, with lower values of $\rho_h$ than in the "26" and "46" designs for $\phi_h < 1$. Indeed, the soft phases in the $P_{24}$ and $G_{24}$ designs are always larger for a given value of $\phi_h$ (provided $\phi_h < 1$) than in the other designs of the same family (Figure 2c-d). Since these plots were made at constant shell thickness ($t$), these relations also approximate the scaling of the surface area with $\phi_h$ ($S_{solid} \approx \frac{V_{solid}}{t}$). Hence, for a fixed value of $\phi_h$, these plots indicate that the total surface area is lower in the "24" designs. This is because the "24" skeleton designs do not contain the locally flat mesh regions (Figure 3a), which are the largest contributors to the overall unit cell area. Specifically, if the Enneper-Weierstrass map is applied to a uniformly-meshed integration domain with equal-area triangles, then the corresponding fundamental patch triangles in the regions with small Gaussian curvature will have larger area than the triangles in regions with strongly negative Gaussian curvature (Figure 2a). Indeed, the area element of the P and G minimal surfaces at a local point is inversely related to the Gaussian curvature at that point (Materials & Methods).

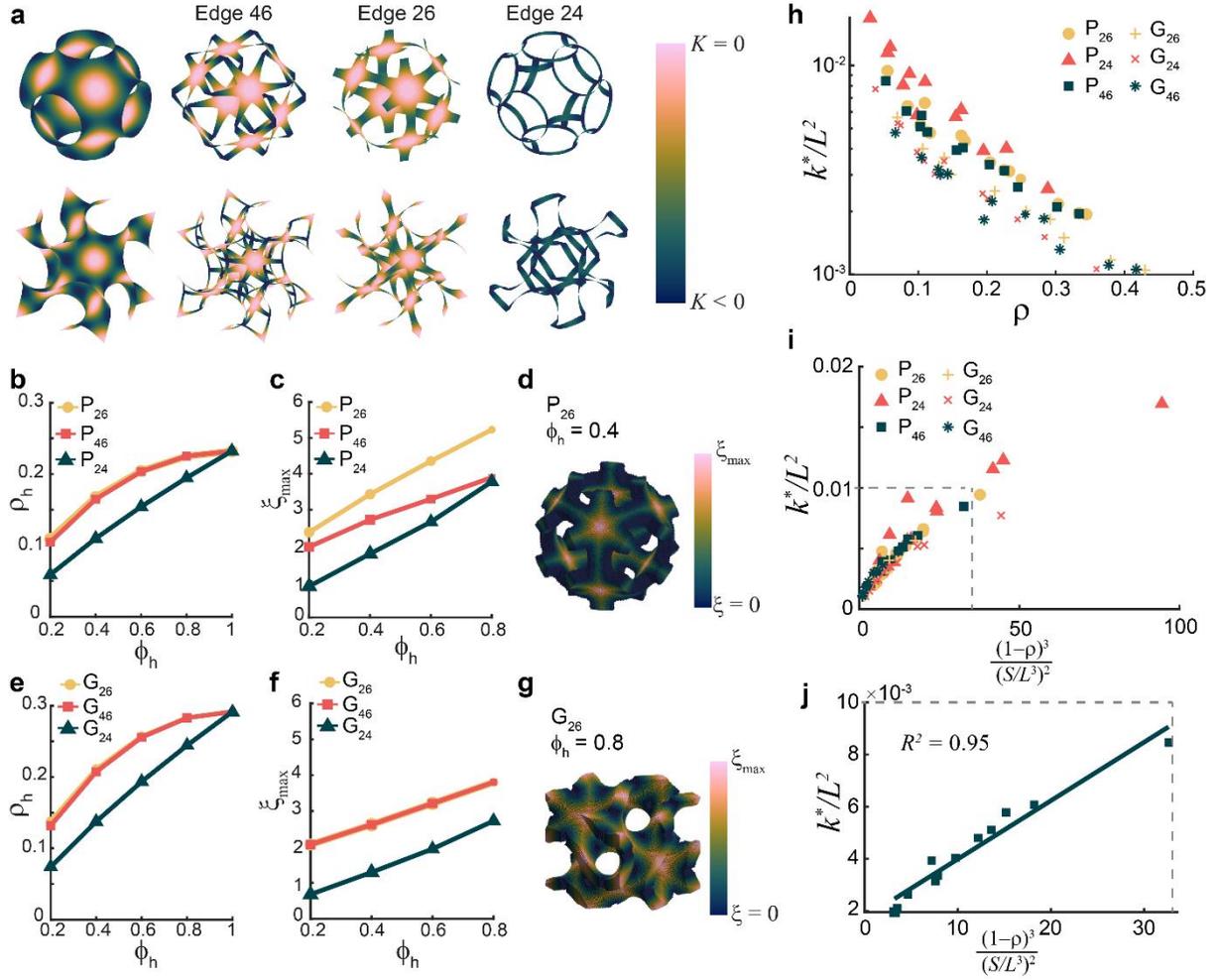

**Figure 3: Morphology and permeability of TPMS-based metamaterials.** a) The Gaussian curvature distribution of the P and G surfaces, as well as their different skeletonized representations. b) The hard-phase volume fraction ($\rho_h$) versus the offset parameter ($\phi_h$) for the different P surface designs with a shell thickness of $t = \frac{L}{10}$. c) The maximum shell factor ($\xi_{max}$) versus the offset parameter ($\phi_h$) for the different P surface designs with a shell thickness of $t = \frac{L}{10}$. d) The visualization of the shell factor $\xi$ for the $P_{26}$ design with $\phi_h = 0.4$ and $t = \frac{L}{10}$. e-f) Analogous to the plots shown in b-c) but for the G surface designs. g) The visualization of $\xi$ for the $G_{26}$ design with $\phi_h = 0.8$ and $t = \frac{L}{10}$. h) The normalized effective permeability ($k^*/L^2$) versus volume fraction ($\rho$) for full and skeleton P and G designs. i) $k^*/L^2$ versus the geometric factor $\frac{(1-\rho)^3}{(S/L^3)^2}$ for all the designs in h). j) A magnified view of the data for the $P_{46}$ designs in i).

In order to study the local shell-like or strut-like nature of the hard phase, which forms the reinforcing backbone of the entire unit cell, we introduced the shell factor $\xi$. We defined $\xi$ for any point in the hard phase as the shortest distance to the soft phase, divided by the shell thickness of the unit cell (Materials & Methods). As such, $\xi$ is a measure of the largest circular shell that locally fits inside the hard phase at every point, with larger $\xi$ representing a locally more shell-like morphology. We quantified $\xi_{max}$ for the P (Figure 3c-d) and G (Figure 3f-g) designs as a function of the offset factor $\phi_h$. All the three P designs exhibited a different scaling of $\xi_{max}$ with $\phi_h$, while the $G_{26}$ and $G_{46}$ showed the same scaling behavior. Moreover, the $P_{24}$ and $G_{24}$ designs achieved the lowest values for $\xi_{max}$, indicating a more strut-like morphology across the range of $\phi_h$. This is the consequence of the selective removal of flat regions in these designs (Figure 3a).

We were interested in the fluid permeability of the different metamaterial designs, as this is an important property in various applications. In meta-biomaterials, for example, permeability affects the supply of nutrients and oxygen to cells, the ingrowth of regenerated tissue, and the potential biodegradation behavior of the scaffolds [24]. Therefore, we estimated the effective intrinsic permeability using a lattice-Boltzmann simulation scheme (Materials & Methods). The permeability is entirely determined by the unit cell geometry, and is independent of the shape and size of the partitioned domains. For example, all P designs in Figure 2c would exhibit the same permeability as their overall geometry is that of the standard P surface. Therefore, the only parameters affecting permeability are the design type, the shell thickness, and the total offset parameter $\phi = \phi_h + \phi_s$. As expected, the normalized effective permeability $k^*/L^2$ decreases with the volume fraction $\rho$ (Figure 3h), following a similar trend for all designs. Moreover, the permeabilities of the designs based on the G surface are consistently lower than those of the P surface designs. This has previously been observed for full P and G designs, and was attributed to the lower specific surface area of the P surface [25]. While permeability clearly scales inversely (and nonlinearly) with the volume fraction, it is not the only geometric parameter of relevance. We find that the permeability values scale almost linearly with $\frac{(1-\rho)^3}{(S/L^3)^2}$, where $S$ is the surface area of the unit cell and $L$ is the bounding box length (Figure 3i-j). This factor also appears in the so-called Kozeny equation for predicting the permeability of porous materials, and indicates that the specific surface area ($S/L^3$) also plays a role in dictating the metamaterial permeability [24-25].

## 2.4. Elastic mechanical properties

The other set of key properties of interest in this study are the elastic mechanical properties, which depend not only on the unit cell geometry (*i.e.*, $\rho$, $\phi$, and $t$) but also on the bulk properties of both materials. We computed the effective elastic properties of the resulting designs using a computational homogenization scheme (Materials & Methods). Through this finite element-based approach, we calculated the effective stiffness tensor $\boldsymbol{C}^*$ for every design, from which properties, such as the effective elastic modulus ($E^*$), bulk modulus ($K^*$), shear modulus ($G^*$) and Poisson's ratio ($\nu$) could be obtained (Materials & Methods).

The effective elastic modulus $E_{11}^*$, i.e. the stiffness under uniaxial loading in the $\langle 100 \rangle$ direction, of the uniphasic skeleton and full structures scaled according to a power law of $\rho$ (Figure 4a), as expected from the well-known Gibson-Ashby relationships [26]. The weakest structures corresponded to the strut-like $P_{24}$ and $G_{24}$ designs with an offset parameter of $\phi = 0.2$ and a shell thickness of $t = \frac{L}{20}$. All of our strut-like designs (i.e., those with low $\phi$), correspond to bending-dominated architectures according to the Maxwell-Calladine criterion, indicating sub-optimal stiffness [27]. The stiffest structures, corresponding to G-based shell-like architectures ($\phi \geq 0.8$), achieved specific stiffness values close to the Hashin-Shtrikman upper bound (HSU) for nearly isotropic structures, which was also the case for the bulk and shear moduli (Supplementary Figure 1a-b) [5b]. It is, however, important to realize that the priority in (and the novelty of) this study is achieving high tunability in mechanical and mass transport properties, rather than presenting new geometries that achieve extreme (specific) properties. Finally, we also observed positive effective Poisson's ratios for all designs (Supplementary Figure 1c).

The central concept of our design approach is the ability to parametrically partition the unit cells into two distinct domains. As such, we are able to tune the mechanical properties using a combination of geometry and material distribution. To study the effects of material choice, we plotted the effective stiffness against the ratio of the Young's moduli of the hard and soft materials ($E_h/E_s$) for all six design types, with $\phi_h \in [0.2, 0.8]$ and $\phi_s = 1 - \phi_h$ (Figure 4b). When $E_h/E_s = 1$, the behavior of the standard uniphasic P or G unit cell is obtained. For increasing values of $E_h/E_s$, the stiffness reduces for all six design types. As expected, this stiffness reduction is much stronger for the lower values of $\phi_h$, where the proportion of the hard phase is low. For example, the stiffness of the $P_{26}$ design with $\phi_h = 0.8$ and $E_h/E_s = 10^3$ is 15% lower than at $E_h/E_s = 1$, while it is 92% lower when $\phi_h = 0.2$ (Figure 4b). Moreover, in all designs, the stiffness reduction curves flatten out when $E_h/E_s \approx 100$. Beyond this point, the soft phase hardly contributes to the overall stiffness, and the load is primarily carried by the hard phase, which forms a reinforcing skeleton for the overall unit cell. Additionally, the stiffness reduction behavior varies among the different designs. For the same values of $\phi_h$, the $P_{24}$ and $P_{46}$ designs exhibit the highest and lowest reduction behavior, respectively. This could be attributed to the overall geometry of the hard phase in both cases: the $P_{24}$ skeleton consists entirely out of highly-curved, slender struts, while the $P_{46}$ skeleton contains a higher number of struts, which are approximately straight and, hence, more efficiently carry load (Figure 3a).

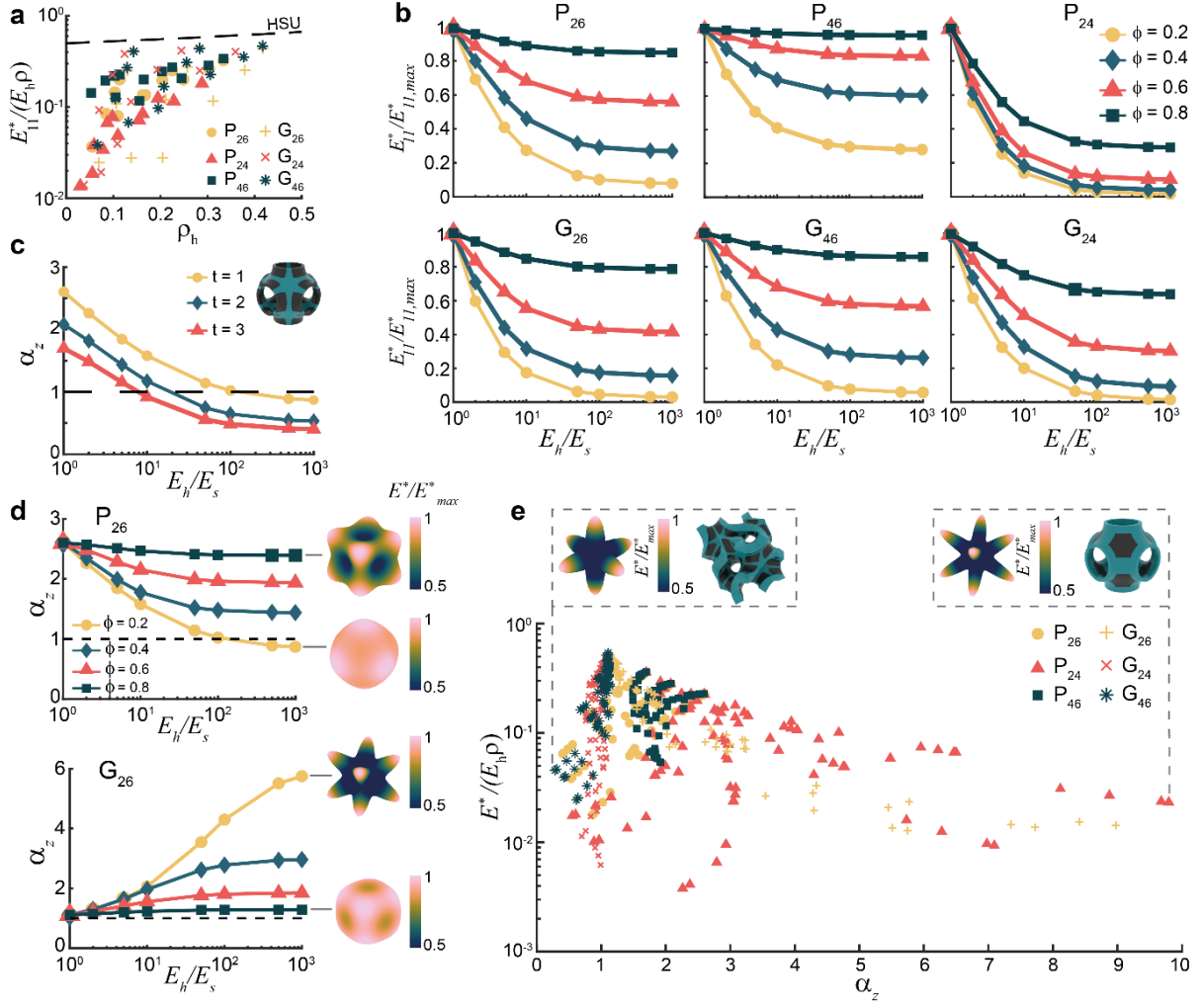

**Figure 4: The elastic mechanical properties and anisotropy of TPMS-based metamaterials.** a) The normalized effective elastic modulus for uniphasic full and skeleton designs ($\phi_s = 0$). The HSU-line indicates the Hashin-Shtrikman upper bound. b) The effective stiffness versus the ratio of the Young's moduli of both phases ($E_h/E_s$) for the six biphasic design types with $t = L/20$ and $\phi_s = 1 - \phi_h$. c) The Zener anisotropy index ($\alpha_Z$) versus $E_h/E_s$ for the $P_{24}$ design with different thicknesses. d) Some examples of the scaling of $\alpha_Z$ with $E_h/E_s$ for the different amounts of the hard phase ($\phi_h$) in the $P_{26}$ (top) and $G_{26}$ (bottom) designs. The inset figures on the right represent the effective elastic surface (effective modulus in all directions). e) The normalized effective stiffness versus Zener anisotropy index. The inset figures show the elastic surfaces and unit cells for both extremal designs regarding $\alpha_Z$.

The elastic properties of metamaterial architectures are, in general, anisotropic, though isotropic variants have been proposed [5a]. We quantified the elastic anisotropy of the different designs as function of $E_h/E_s$ using the Zener anisotropy index $\alpha_Z$ (Materials & Methods). When $\alpha_Z = 1$, the structure is elastically isotropic, meaning that the effective stiffness is equal in all directions. We found that $\alpha_Z$ varies with $E_h/E_s$ and shell thickness $t$ for all the designs, although the extent to which it varies depends on the design type (Figure 4c-d and Supplementary Figure 1). For example, the standard, uniphasic unit cell of the P designs (obtained at $E_h/E_s = 1$) is anisotropic, with $\alpha_Z > 1$. However, the anisotropy index changes with $E_h/E_s$ for biphasic designs, as exemplified in the $P_{26}$ design (Figure 4d). For $\phi_h = 0.8$, $\alpha_Z$ remains almost constant throughout the range of material ratios. In this case, the effective elastic surface, representing the effective stiffness in all directions (Figure 4d), indicates a higher stiffness in the $\langle 111 \rangle$ direction as opposed to the $\langle 100 \rangle$ direction. However, for $\phi_h = 0.2$, $\alpha_Z$ rapidly reduces with increasing $E_h/E_s$, achieving an isotropic design when $E_h/E_s \approx 10^2$. The G designs, on the other hand, start of as quasi-isotropic structures for $E_h/E_s = 1$. For the $G_{26}$ design specifically, an increase in $\alpha_Z$ was observed as $E_h/E_s$ increased, in particular when $\phi_h = 0.2$ (Figure 4d). To summarize the elastic property space, we plotted the normalized effective modulus versus the Zener anisotropy index (Figure 4e). We found that most anisotropic designs have $\alpha_Z > 1$, although some designs (*e.g.*, for $G_{46}$ and $P_{26}$) exhibited $\alpha_Z < 1$. The largest spread in anisotropies were found in the $G_{26}$ and $P_{24}$ designs, the latter reaching a maximum of $\alpha_Z \approx 10$. Moreover, we observed a relatively large number of designs in the quasi-isotropic range ($\alpha_Z \approx 1$), which cover approximately two orders of magnitude in the effective stiffness. Overall, these results confirm that our metamaterial design strategy not only enables the effective tuning of the uniaxial stiffness, but also of the elastic anisotropy.

**2.5. Balancing elasticity and permeability**

Our parametric design approach substantially enhanced the ability to independently tailor the mechanical and mass transport properties of TPMS-based metamaterials. This is visualized in the elasticity-permeability property space, where the effective elastic modulus normalized by the hard-phase elastic modulus ($E^*/E_h$) is plotted against the intrinsic (area-normalized) effective permeability ($k^*/L^2$) for all uni- and biphasic designs (Figure 5a). In the case of the "full" unit cells (*i.e.*, the unit cells with the standard P or G morphology, $\phi_s = 1 - \phi_h$), the biphasic partitioning unlocks a wide range of attainable stiffness values for a constant value of permeability (the data points in the yellow bands in Figure 5a). Indeed, for those designs, the permeability is only determined by the overall unit cell type (P or G) and shell thickness ($t$), while the stiffness is also driven by the material distributions. The maximum stiffness for the full designs is, not surprisingly, obtained for $\phi_h = 1$ (*i.e.*, unit cells that consist entirely out of the hard phase).

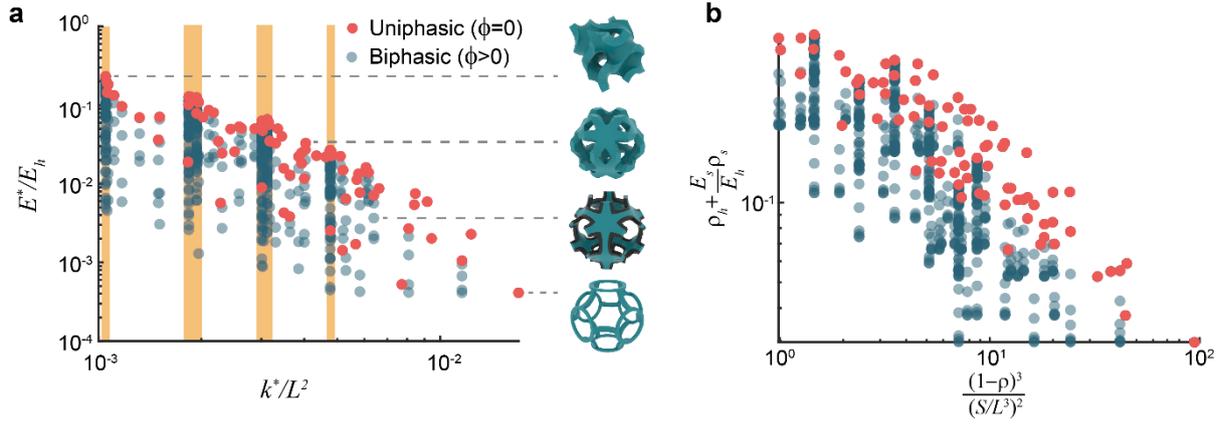

**Figure 5: Effective stiffness versus effective permeability.** a) The normalized effective elastic modulus versus the normalized effective permeability for uni- and biphasic designs. The yellow bands indicate the "full" designs (*i.e.*, $\phi_s + \phi_h = 1$). The unit cell structures belonging to four data points are visualized in the inset figures. b) Simple scaling laws that capture the behavior from a). The elastic properties are determined by the geometry and materials choice, captured by $\rho_h + \frac{E_s}{E_h}\rho_s$ while the permeability depends on the geometry alone, as described by $(1-\rho)^3/(S/L^3)^2$.

While biphasic partitioning enables continuous stiffness tuning in standard TPMS unit cells, their intrinsic permeability range is still limited and is only a function of the shell thickness ($t$). However, the range of intrinsic permeability values is extended by the skeleton TPMS structures ($\phi_h + \phi_s < 1$), which have more holes in their surfaces, thereby altering the fluid flow through the unit cell. Indeed, all data points outside of the yellow bands in Figure 5a correspond to uni- or biphasic skeleton designs. Similar to the case of full designs, using two different materials in the skeleton designs enables the tuning of the elastic properties independently from permeability. Alternatively, one could fix the effective normalized stiffness and tune the permeability by choosing a different design at the same level of stiffness. We note that we have visualized the elasticity-permeability design space using discrete values of $t$, $\phi_h$, $\phi_s$ and $E_h/E_s$ in Figure 5, and that intermediate data points could be obtained at intermediate values of the design parameters.

Our design approach enables this level of mechanical and mass transport tunability by leveraging the fact that permeability solely depends on geometry, while the elastic properties depend on geometry and material choice. To further demonstrate this, we plotted $\rho_h + \frac{E_s}{E_h}\rho_s$ against $\frac{(1-\rho)^3}{(S/L^3)^2}$ (Figure 5b), which captures the same trend as in the elasticity-permeability map. The former quantity captures the combined effect of the geometry and material properties on the effective stiffness: the stiffness is primarily determined by the volume fraction of the hard phase, while the soft phase has a weighted contribution, depending on its stiffness relative to the hard phase. The second quantity ($\frac{(1-\rho)^3}{(S/L^3)^2}$) is the purely geometry-dependent metric that was introduced before (Figure 3i-j) and that correlates with the permeability. Taken together, these results summarize our enhanced ability to tune elasticity and permeability independently, by combining spatial distribution of multiple materials with geometric control over the unit cell architecture.

**2.6. Multi-material additive manufacturing**

We physically realized metamaterial lattices based on the P surface, using multi-material additive manufacturing (Materials & Methods). The lattices consisted of 27 unit cells, in a 3×3×3 arrangement (Figure 6a). We printed two uniphasic full designs in a hard (top left in Figure 6a) and soft (bottom right in Figure 6a) polymer, as well as a uniphasic $P_{46}$ skeleton design ($\phi_h = 0.2$, $\phi_s = 0$, top right in Figure 6a), and a biphasic $P_{46}$ design ($\phi_h = 0.2$, $\phi_s = 0.8$, bottom left in Figure 6a). All structures were successfully printed, and showed no signs of defects. In the case of the biphasic design, the soft phase was well integrated with the hard-phased skeleton.

We mechanically tested the specimens by imposing a small macroscale compressive strain (Figure 6b and Materials & Methods), in order to examine the effective elastic modulus. Compared to the fully hard structure, the stiffness of the skeletonized structures (uni- and biphasic) was 70-80% lower, while the stiffness of the fully soft structures was three orders of magnitude lower (Figure 6c-d). We observed that the stiffness reduction in the biphasic design corresponded well with the computationally predicted stiffness reduction (dotted line in Figure 6c). However, the stiffness of the uniphasic skeleton was higher than that of the corresponding biphasic design, while their predicted stiffness values are equal. While this difference warrants further investigation, we believe that is the consequence of differences in hard-phase thickness due the presence or lack of the soft phase. For example, the transition region of the hard and soft phase in the biphasic design might locally have different material properties than in the uniphasic skeleton design, due to the localized mixing of the polymer droplets in the material jetting process. Nevertheless, these lattices demonstrate that the biphasic TPMS-based metamaterials can be successfully manufactured using commercially available printing processes.

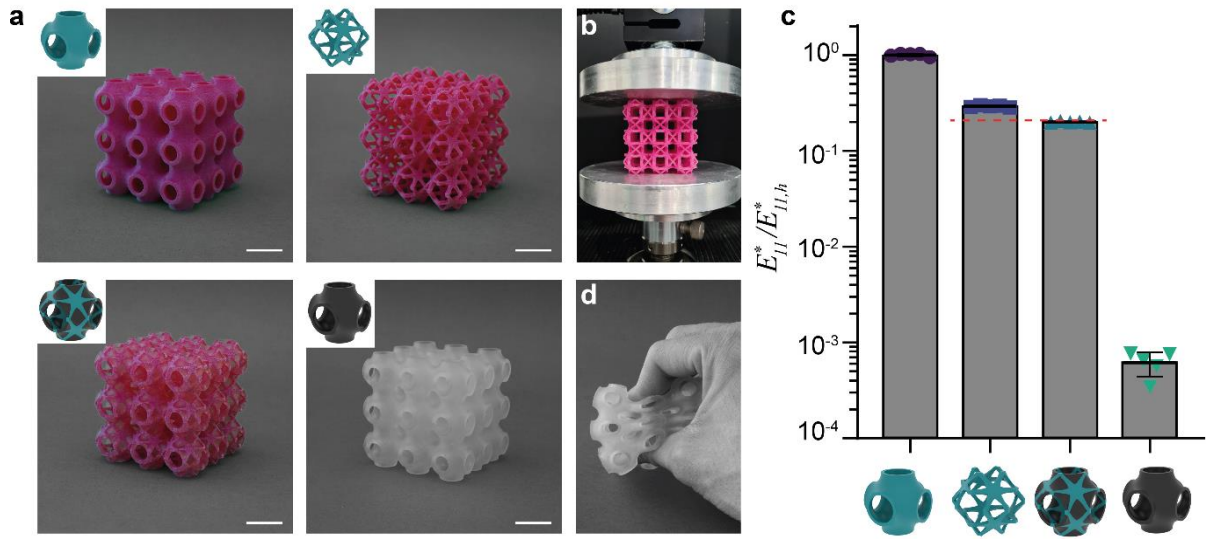

**Figure 6: Additively manufactured structures and mechanical testing.** a) Lattices consisting of 27 unit cells fabricated using multi-material polymer printing. The pink material is a stiff polymer (VeroMagenta), while the white transparent material is soft (Agilus30) and easily deformable. b) The mechanical testing of the 3D-printed structures using a displacement-controlled compression test setup. c) Normalized compression testing results for the four printed specimens, showing the uniaxial effective modulus. Each test was repeated five times. The markers show the individual test results, while the bar graph represents the mean and standard deviations. The red dashed line corresponds to the predicted stiffness for the uni- and biphasic skeletons. d) The structure that consists entirely out of the soft phase is highly deformable and has a negligible stiffness as compared to the other designs.

## 3. DISCUSSION

We have presented an approach to parametrically design multifunctional metamaterials in which the mechanical and mass transport properties can be decoupled to a large extent. The ability to independently tailor such properties is relevant in many applications, yet is challenging to achieve in uniphasic metamaterials due to the competing dependence of both properties on the metamaterial geometry [12d]. Our design approach relies on partitioning the P and G TPMS unit cells into hard, soft, or void domains, according to templates that are based on hyperbolic tilings, effectively resulting in hybrid strut-shell-based metamaterials. The permeability of the structures is purely geometry-driven and can be tailored by changing the shell thickness or through the controlled introduction of additional openings in unit cells. The elastic properties, on the other hand, also depend on the choice of the materials and their spatial distribution. We showed that these biphasic decorations enabled us to achieve a wide range of effective stiffness values at fixed permeability, but also offered a route to tailor the elastic anisotropy (maintaining cubic symmetry). While we have focused on elastic properties here, it is likely that the combination of hard and soft phases would also affect the other mechanical properties of TPMS-based architectures, such as the energy absorption at high strains, crack growth (the soft inclusions might act as crack inhibitors), or fatigue response. In any case, our results confirm that the ability to control the material architecture of multiple materials instead of a single one can significantly expand the overall property space of the resulting [17b, 17c] and enabling the multi-objective design of multi-physics metamaterials.

Our approaches could be generalized and extended in various ways. First, it would be possible to vary the unit cell type throughout the metamaterial lattice, for example, to spatially vary the permeability without affecting the local elastic properties. This could be achieved by fixing the geometry of the hard phase in all the unit cells, but varying the amount of the soft phase (Supplementary Figure 2a). Alternatively, the overall unit cell geometry could be preserved, hence fixing permeability, but the elasticity could be spatially tuned by varying the amount of the load-carrying hard phase (Supplementary Figure 2b). Our approach could also be extended to enable other types of shell-based biphasic metamaterials. For example, the same hyperbolic tilings that we have used here could be projected onto the D (diamond) minimal surface, which belongs to the same family as the P and G surfaces (Supplementary Figure 3). We did not include the D designs in this study, as these unit cells contain non-manifold regions where different patches meet along single edges, rendering them unattractive from a mechanical viewpoint. Moreover, many other biphasic P and G designs could be created, beyond the ones we have presented here. For example, the tilings that we have used could be combined together to form hybrid structures (Supplementary Figure 4). Alternatively, many other tilings exist that could be projected onto these cubic TPMS to make different admissible biphasic designs [19a, 19c]. In addition to the PDG surface family of minimal surfaces, our design approach could be extended to other TPMS families, provided their Enneper-Weierstrass parametrization is known. The Weierstrass function has already been determined for several other TPMS families, but can also be uncovered for newer types of TPMS on the basis of the local flat points, which could be determined numerically [28]. Finally, the central concept of decorating a lattice with two (or more) different materials could also be applied to stochastic microstructures. Recently, such stochastic shell-based architectures have emerged as attractive metamaterial geometries, due to their high tunability, large design space, and robustness against deteriorating symmetry-breaking defects [6d, 29]. However, applying the multi-material methodology to such stochastic geometries would require a different strategy to rationally distribute the different phases, for example, by parametrically applying a skeletonization algorithm to obtain the medial graph (or an inflated version) of the shell-based structure [30], and assigning different material properties to this region than to the remainder of the geometry.

From a broader perspective, this study underscores the relevance of reticular (or structural) chemistry as a source of inspiration for metamaterial design. In addition to the hyperbolic networks described here, there are vast databases of complex topologies that could be used as metamaterial templates [20]. For example, the wealth of zeolitic networks and metal-organic frameworks (MOFs) could inspire complex designs that go beyond the traditional lattice choices (*e.g.* cubic or diamond lattices), resulting in so-called "meta-MOFs" [31]. Moreover, metamaterial properties could be enhanced even further by incorporating the same hardening mechanisms that are found at the atomic scale in crystalline materials [32].

Taken together, we demonstrated a new strategy for the multi-objective design of multi-physics metamaterials and a route to decouple properties that are conflicting in uniphasic metamaterials. By leveraging the hyperbolic symmetries of TPMS, our design approach maintains a surprising tractability, yet produces complex 3D, biphasic architectures. Together with advances in multi-material additive manufacturing, this design approach could unlock exciting routes towards multi-physics metamaterials in a variety of applications. In this regard, we believe that decoupling elasticity and permeability is only one example of the potential for enhanced tunability in biphasic metamaterials.

**ACKNOWLEDGEMENTS**

S.J.P.C. is grateful to Dr. Kevin M. Moerman for helpful discussions and guidance in the use and adaption of the functions in the GIBBON toolbox. We are grateful to 3D LifePrints (Oxford, UK) for 3D-printing the specimens shown in Figure 6. The research leading to these results has received funding from the European Research Council under the ERC grant agreement no. [677575].

## 4. MATERIALS & METHODS

### 4.1. Parametric design of TPMS

The 3D mesh representations of the labelled P and G surfaces were computed using the Enneper-Weierstrass parametrization, which maps the points in an integration domain in the complex plane to the curved fundamental patch in $\mathbb{E}^3$ that is used to build the TPMS. Specifically, the Cartesian coordinates of the points on the fundamental patch are obtained by:

$$\begin{Bmatrix} x \\ y \\ z \end{Bmatrix} = Re\left[ e^{i\theta} \int_{\omega_0}^{\omega} \begin{Bmatrix} 1-\widetilde{\omega}^2 \\ i(1+\widetilde{\omega}^2) \\ 2\widetilde{\omega} \end{Bmatrix} R(\widetilde{\omega})d\widetilde{\omega} \right] + p_0$$

Here, $R(\widetilde{\omega})$ is the Weierstrass function, $\theta$ is the Bonnet angle, $p_0$ is an arbitrary translation to define the origin (here, $p_0 = [0,0,0]$), $\omega_0$ is a fixed point in the integration domain (here, $\omega_0 = 0$), and $\omega$ is any other point in the integration domain. For the PDG surface family, the Weierstrass function is defined as:

$$R(\widetilde{\omega}) = [\widetilde{\omega}^8 - 14\widetilde{\omega}^4 + 1]^{-\frac{1}{2}}$$

For the P surface, $\theta = \frac{\pi}{2}$. For the G surface, $\theta = \text{arccot}\left(\frac{E_k\left(\frac{3}{4}\right)}{E_k\left(\frac{1}{4}\right)}\right)$, where $E_k(k)$ is the complete elliptic integral of the first kind with parameter $k$. Thus, any point $\omega$ in the complex domain is mapped to a point in the 3D fundamental patch through this parametrization. The points $\omega$ were uniformly sampled from the complex domain, depending on the patch type and the desired density. In case of the $P_{46}$ design, for example, all points were sampled from parallel lines to the "46" edge of the complex domain. Next, the Delaunay triangulation of the set of discrete points $\omega$ was computed to obtain a triangular (2D) mesh of the complex domain (Figure 2a). The faces of the 2D mesh were labelled as hard, soft, or void phase (in parallel bands), depending on the desired offset parameters $\phi_h$ and $\phi_s$ (Figure 2a). The Enneper-Weierstrass equations were then used to map the points $\omega$ to their Cartesian coordinates in $\mathbb{E}^3$. The mesh topology and face labelling that was computed on the 2D complex domain was applied to the 3D set of points to obtain a meshed representation of the fundamental patch. Finally, the patch was patterned in 3D according to the P and G symmetry operations to obtain the translational unit cells [22].

All computations and consequent mesh processing steps were performed in MATLAB (MATLAB 2018b, Mathworks, Natick, MA, USA) using custom code, as well as by using several of the functions of the GIBBON toolbox [33].

### 4.2. Conversion to solid structures

The zero-thickness meshes were converted to solid, 3D-printable structures by a surface thickening approach. To this end, all vertices were offset in the positive and negative normal directions by a distance $d/2$, where $d$ is a user-defined fraction of the unit cell bounding box length $L$ (*i.e.*, $d = \frac{t}{20}L$). This offsetting operation resulted in two parallel meshes, one at each side of the original minimal surface mesh. Triangular bounding faces were added at the edges of the two parallel meshes to create a watertight mesh that represents the solid structure. This thickening approach was applied for every labeled region of the mesh separately, resulting in a solid triangle mesh for both the hard and soft phases. The surface area and relative densities were then computed on the basis of these triangle meshes, and the meshes were exported in the STL format for printing and visualization in Keyshot (Keyshot 5, Luxion, Tustin, CA, USA). The voxelized representations of the unit cells were created from the triangle meshes using the function *patch2Im* in the GIBBON toolbox [33].

### 4.3. Morphology

The Gaussian curvature ($K$) of the TPMS mesh vertices (Figure 3a) was computed from the complex domain as [34]:

$$K(\omega) = -4(1 + |\omega|^2)^{-4}|R(\omega)|^{-2}$$

The area element ($dS$) or surface metric for a point $\omega = u + iv$ is defined as [35]:

$$dS = dudv(1 + |\omega|^2)^2|R(\omega)|^2$$

The shell factor $\xi$ (Figure 3c-d and Figure 3f-g) was computed on the voxelized mesh representations (100×100×100 voxels) of the entire unit cell. First, the soft phase of the unit cells was thresholded, resulting in a binary 100×100×100 array with label 1 for all the voxels in the soft phase, and label 0 for all other voxels. Then, the Euclidean distance map (EDM) was computed on this 3D array, specifying the distance from every voxel to the nearest voxel in the soft phase. The hard phase was then used as a mask to extract the distance of every voxel in the hard phase to the nearest voxel in the soft phase. The distance value was then divided by the local shell thickness to compute the shell factor at every point $p$ in the hard phase (the factor 2 is to obtain the diameter of the largest circular shell that fits inside the hard phase at every point):

$$\xi(p) = 2 \cdot \frac{EDM(p)}{t(p)}$$

### 4.4. Permeability simulations

The effective fluid permeability for the different uniphasic designs was computed using a lattice-Boltzmann (LB) scheme that has previously been used for determining permeabilities of standard TPMS microstructures [12c]. Briefly, the LB method models the temporal evolution of a particle velocity distribution function at discrete lattice positions under collision and streaming steps, and subject to a small pressure gradient [36]. Here, the LB simulations were performed with standard D3Q19 elements (*i.e.*, 3D elements with 19 possible momentum components) [37] and a lattice discretization of $256^3$ voxels was used. The intrinsic permeability was extracted using the Darcy's law, and were normalized to the cross-sectional area of the unit cell $L^2$.

**4.5. Effective elastic properties**

The effective elastic mechanical properties of the different unit cell designs were computed using a computational homogenization scheme based on the finite element method (FEM) in MATLAB [38]. Briefly, the effective (homogenized) elasticity tensor $\boldsymbol{C}^*$ (6×6, using Voigt notation) was extracted from six independent, linear elastic FEM simulations on voxelized representations of the unit cells with periodic boundary conditions. In the simulations, the Poisson's ratio was set to 0.3, and the stiffness of the hard phase was set to 2 GPa. Following a convergence study on the basis of the effective elastic modulus (considered converged when the variation was below 1%), a discretization of the unit cells into $128^3$ voxels was found to be sufficient to compute the effective elastic properties. The linear force-displacement equations in the FEM simulations were solved using the preconditioned conjugate gradient (pcg) scheme in MATLAB, with a tolerance set to $10^{-8}$. Due to the cubic symmetry of the metamaterial designs, $\boldsymbol{C}^*$ contains only three independent components (*i.e.*, $C_{11}$, $C_{12}$ and $C_{44}$). From those components, the effective uniaxial Young's modulus ($E_{11}^*$ in $\langle 100 \rangle$ direction), effective bulk modulus ($K^*$), effective shear modulus ($G^*$, *e.g.* applied on the (100) plane in the [010] direction), and the effective Poisson's ratio ($\nu^*$, for loading in the $\langle 100 \rangle$ direction) could be determined [7d] as:

$$E_{11}^* = \frac{1}{S_{11}^*}$$

$$K^* = \frac{C_{11} + 2C_{12}}{3}$$

$$G^* = C_{44}$$

$$\nu^* = \frac{C_{12}}{C_{11} + C_{12}}$$

Here, $S_{11}^*$ is the (1,1) component of the homogenized compliance tensor $S^*$. The Zener anisotropy index for cubic crystals ($\alpha_Z$) was determined as [7d]:

$$\alpha_Z = \frac{2C_{44}}{C_{11} - C_{12}}$$

To plot the elastic surfaces (Figure 4d-e), the effective Young's modulus in different directions was calculated by transforming the effective stiffness tensor using the appropriate rotation matrix for every direction [38].

The Hashin-Shtrikman (HS) upper bounds for the effective bulk ($K_{HSU}$) and shear ($G_{HSU}$) moduli of a nearly isotropic material were computed as [5b]:

$$K_{HSU} = \frac{4G_s K_s \rho}{4G_s + 3K_s(1-\rho)}$$

$$G_{HSU} = \frac{(9K_s + 8G_s)\rho G_s}{20G_s + 15K_s - 6(K_s + 2G_s)\rho}$$

The corresponding HS bound for the Young's modulus (assuming isotropic linear elasticity) was then determined as a function of $K_{HSU}$ and $G_{HSU}$ is given by:

$$E_{HSU} = \frac{9K_{HSU} G_{HSU}}{3K_{HSU} + G_{HSU}}$$

### 4.6. Multi-material additive manufacturing

Four different designs were additively manufactured through a material jetting process, using a combination of hard and soft photocurable polymer resins. The length of the unit cell bounding box was set to $L = 20$ mm, the shell thickness was $t = 2$ mm, and the lattices consisted of 27 unit cells in a 3×3×3 arrangement. The fabrication was performed using a Connex3 Objet 350 printer (Stratasys, Minnesota, USA). Both the hard and soft phases were made with commercial polymer resins designed for this printing system: the hard phase was printed using the VeroMagenta polymer (Stratasys, $E$=2-3 GPa according to the manufacturer), while the soft phase was printed using the translucent, rubber-like Agilus30 polymer (Stratasys, Shore A hardness: 30-35 according to the manufacturer, which corresponds to $E \approx 1.2$-1.4 MPa using Gent's relation [39]). This combination of materials has previously been used to print 2D metamaterials with $E_h/E_s \approx 10^3$ [17c]. The lattices were printed with soluble support material (SUP706, Stratasys), which was carefully removed after printing using chemical washing (according to the manufacturer's protocol), water rinsing, and compressed air blowing.

### 4.7. Mechanical compression tests

The specimens were mechanically tested in a displacement-controlled uniaxial compression test, using a Lloyd universal test bench (LR5K, Ametek STC, Bognor Regis, UK), equipped with a 5kN load cell. The test was performed under ambient temperatures (22 °C) at a strain rate of $10^{-3}$ s$^{-1}$. A preload of 5 N was used and the test was halted at 2% macroscopic strain.

Every specimen was tested five times, allowing ample time between consecutive tests for the material to recover its original shape. Throughout the low-strain testing, the specimens maintained their integrity and the force-displacement curves did not show signs of failure. The effective Young's modulus was determined from the linear-elastic gradient of the linear part of the stress-strain curve.